\documentstyle[twocolumn,psfig,aps]{revtex}
\begin{document}
\def\v{\rule{.4mm}{2.4mm}}
\def\h{\rule[1mm]{2.4mm}{.4mm}}
\def\half{\frac{1}{2}}
\draft
\title{The Anti-Jahn-Teller Polaron in LaMnO$_3$}
\author{Philip B. Allen and Vasili Perebeinos}
\address{Department of Physics and Astronomy, State University of New York,
Stony Brook, New York 11794-3800}
\date{\today}
\maketitle
\begin{abstract}

Distortions of the oxygen sublattice couple to $e_g$ orbitals
of Mn$^{3+}$ and drive a cooperative Jahn-Teller 
(orbital ordering) transition in LaMnO$_3$.  A simple model
for this transition is studied.  Without further adjustment,
the model predicts the shape and stability
of small (anti-Jahn-Teller) polarons which form
when holes are doped into the material.  This leads to a new description
of the lightly doped insulator, the anti-ferromagnetic to ferromagnetic
transition, and the metal-insulator transition.

\end{abstract}

\section{introduction}

The doped manganites R$_{1-x}$S$_{x}$MnO$_3$ (where R is
typically La and S is typically Sr or Ca) have a fascinating
($T$,$x$) (temperature, concentration) phase diagram, including
``colossal magnetoresistance'' \cite{giant} when $T\approx 250$K and
$x\approx$0.20.  Experiment (transport \cite{transport},
optical \cite{optical}, diffraction \cite{diffraction},EXAFS \cite{exafs},
isotope studies \cite{isotope})
and theory \cite{theory} indicate the appearance of
polarons at this point in ($T$,$x$).  It is less well appreciated that
small polarons are essential to explain the insulating phase at smaller
$x$ and 0K$<T<$750K \cite{Yamada}.  We study a simple model for the cooperative 
Jahn-Teller (JT) transition, and use the model to predict properties of
small polarons in lightly-doped material, including how they affect
magnetic order and the metal-insulator transition.  By working in the
limit of large on-site Coulomb repulsion $U$, we find that properties
of the polaron are simple enough to describe analytically, with
small perturbative corrections.  

There is disagreement over the relative role of Coulomb, magnetic,
and electron-phonon effects.  We offer a simple unified
picture in which the relevant energy scales in descending
order are: (1) Coulomb interactions
are inactive after establishing the dominant Hubbard and Hund energy
scales; (2) electron-phonon interactions drive orbital ordering by the
JT mechanism; (3) when doped, electron-phonon interactions 
{\sl via} orbital ordering
give small anti-JT polarons; and finally (4)
orbital ordering disrupted by polarons is responsible for the unusual 
magnetic phases (described in Sec. III.)

A canonical ``Jahn-Teller polaron'' \cite{Thomas} is
the excess electron in BaTiO$_3$ \cite{Schirmer},
which sits in a triply degenerate $t_{2g}$ level.
A local distortion of the lattice splits
the degenerate levels.  Beyond a critical coupling strength, this
lowers the energy and traps the electron in a small polaron state.

\begin{figure}
\centerline{
\psfig{figure=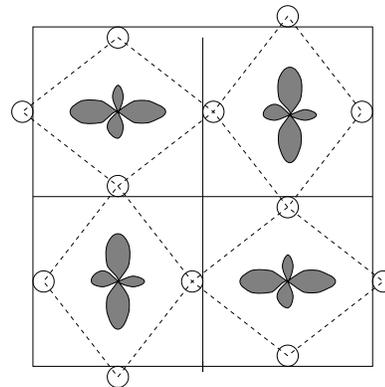,height=2.0in,width=2.0in,angle=0}}
\caption{Base plane (x-y plane) of Jahn-Teller distorted LaMnO$_3$.
Rotations of oxygen octahedra are omitted, and distortions
are exaggerated.  Oxygens are displaced only along Mn-O-Mn bonds. }
\label{jt}
\end{figure}

By contrast, the polarons in LaMnO$_3$ are ``anti-JT polarons.'' 
When pure, the valence is Mn$^{3+}$, with $d^4$ configuration in the high-spin
state $t_{2g\uparrow}^{3} e_{g\uparrow}^{1}$ with the doubly-degenerate
$e_g$ level singly occupied.
Below the JT structural transition at $T_{\rm JT}$=750K \cite{Goodenough},
oxygen octahedra distort as illustrated schematically in Fig. \ref{jt},
lifting the $e_g$ degeneracy and lowering the energy of the occupied  
orbitals.  When a hole is
added, we show that a small polaron is formed by locally
``undoing'' \cite{Yamada} the JT distortion, pinning the hole onto a
Mn$^{4+}$ site with a filled $t_{2g}$ shell.

\section{model Hamiltonian}

\begin{figure}
\centerline{
\psfig{figure=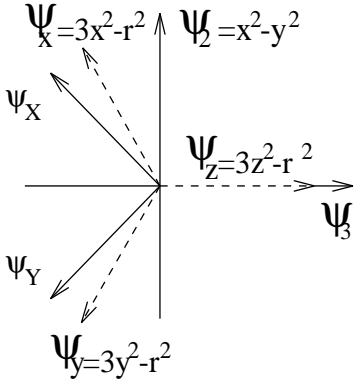,height=2.07in,width=2.0in,angle=0}}
\caption{The two-dimensional space of $e_g$ orbitals, with vertical
and horizontal axes being the usual orthogonal basis functions
$\psi_2$ and $\psi_3$.  The symmetrical overcomplete basis functions
$\psi_x$, $\psi_y$, and $\psi_z$, Eqn. \protect\ref{psixyz}, are
shown as dashed lines, and the symmetrical orthogonal basis
$\psi_X$ and $\psi_Y$, Eqn. \protect\ref{psiXY},
are shown lying at \protect45$^{\circ}$
in the second and third quadrants.
}
\label{egspace}
\end{figure}

The Hamiltonian we use is ${\cal H}= {\cal H}_{t}
+{\cal H}_{\rm ep} +{\cal H}_{\rm L} +{\cal H}_{\rm U}$. The first term 
represents hopping of Mn $e_g$ electrons to nearest
neighbors.  A simple way to derive this term is to introduce an
overcomplete basis
\begin{equation}
\psi_x = 3x^2-r^2, \  \psi_y = 3y^2-r^2, \ \psi_z = 3z^2-r^2, 
\label{psixyz}
\end{equation}
each pointing toward the two nearest Mn neighbors along one of
the Cartesian axes.
Note that $\psi_x + \psi_y + \psi_z =0$.  In the two-dimensional
$e_g$ space, these basis vectors lie at 120$^{\circ}$ to each other,
as shown in Fig. \ref{egspace}.  The usual orthogonal basis,
$\psi_2 = \left(\psi_x - \psi_y \right)/\sqrt{3} $ and $\psi_3 = \psi_z $,
is more complicated because there is no element of the cubic rotation
group which transforms $\psi_2$ into $\psi_3$.
Then for the hopping Hamiltonian we choose: 
\begin{equation}
{\cal H}_{\rm t}=t \sum_{\ell,\pm} \left\{[c^{\dagger}_x(\ell)
c_x(\ell \pm \hat{x})] + [x\rightarrow y] +
[y \rightarrow z] \right\}
\label{ht}
\end{equation}
where $\ell$ numbers manganese sites, and $\ell + \hat{x}$ 
numbers the Mn neighbor to the right.  After
re-expressing ${\cal H}_{\rm t}$ in terms of the orthogonal orbitals
$\psi_2$ and $\psi_3$, we recover the correct nearest-neighbor
two-center Slater-Koster \cite{Slater} model, with overlap integrals
$t=(dd\sigma)$, $(dd\pi)$ not entering due to symmetry, and
$(dd\delta)=0$.  

The only lattice degree of freedom is
oxygen motion along the direction of the bonds to the nearest two Mn atoms,
with a harmonic restoring force $-Ku_x$.  
Variables $Q_x(\ell)=u_x(\ell+\half \hat{x})-u_x(\ell-\half \hat{x})$ 
measure the local 
expansion of oxygens on the $x$-axis around the $\ell$-th
Mn atom.  This expansion lowers the energy $\epsilon_x$ of orbital $\psi_x$
by $\partial \epsilon_x/\partial Q_x=-4g/\sqrt{3}$. We work in adiabatic approximation (oxygen mass $=\infty$).  This gives

\begin{equation}
{\cal H}_{\rm ep}=-\frac{4g}{\sqrt{3}}
 \sum_{\ell} \left\{ [c^{\dagger}_x(\ell)c_x(\ell)Q_x(\ell) ]
+ [x\rightarrow y] +[y \rightarrow z] \right\}
\label{hep}
\end{equation}
\begin{equation}
{\cal H}_{\rm L}=\frac{K}{2}\sum_{\ell}[u_x(\ell+\half \hat{x})^2
+u_y(\ell+\half \hat{y})^2+u_z(\ell+\half \hat{z})^2].
\label{hl}
\end{equation}
When two orbitals are occupied on one site, one has to pay Coulomb energy $U$:
\begin{equation}
{\cal H}_{\rm U}=U\sum_{\ell}c_2^{\dagger}(\ell)c_2(\ell)
c_3^{\dagger}(\ell)c_3(\ell) 
\label{hu}
\end{equation}
We take into account $t_{2g}$ spins, by defining the $c^{\dagger}$ 
operators to create electrons whose spins (because of the
large Hund's rule energy) are parallel to the S=3/2 core
spin.  Hopping (Eqn. \ref{ht}) then operates only 
between adjacent Mn sites with parallel spins.  

This model generalizes to $e_g$
electrons the model of Rice and Sneddon \cite{Rice} for $s$ electrons
in BaBiO$_3$.  The same model, but with longer range forces, was used
by Millis \cite{Millis} for LaMnO$_3$.

To analyze the solution, it is convenient to make a 
45$^{\circ}$ rotation in $(\psi_2,\psi_3)$-space,
to a new orthogonal basis, shown by the solid arrows in Fig. \ref{egspace},

\begin{eqnarray}
\psi_X&=& \ \ \frac{1}{\sqrt{2}}(\psi_2-\psi_3)
       =\frac{1}{\sqrt{6}}\left(
        \left(\sqrt{3}+1\right)\psi_x+\left(\sqrt{3}-1\right)\psi_y\right)
\nonumber \\
\psi_Y&=&-\frac{1}{\sqrt{2}}(\psi_2+\psi_3)
        =\frac{1}{\sqrt{6}}\left(
        \left(\sqrt{3}-1\right)\psi_x+\left(\sqrt{3}+1\right)\psi_y\right)
\label{psiXY}
\end{eqnarray}
The orbitals
$\psi_X$ and $\psi_Y$ point strongly in the $\hat{x}$ and $\hat{y}$
directions, and are equivalent under a 90$^{\circ}$ rotation in
real space, as shown in Fig. \ref{jt}.

In the new basis the electron-phonon term
in the Hamiltonian (\ref{hep}) is conveniently split into Jahn-Teller and
breathing parts, ${\cal H}_{\rm ep}={\cal H}_{\rm JT}+{\cal H}_{\rm br}$:
\begin{equation}
{\cal H}_{\rm JT}=-g\sum_{\ell}\left(c_X^{\dagger}(\ell), 
c_Y^{\dagger}(\ell)\right)
\left( \begin{array}{rr}  Q_2(\ell) & Q_3(\ell) \\
             Q_3(\ell) &  -Q_2(\ell)  \end{array} \right)
\left( \begin{array}{c} c_X(\ell)\\c_Y(\ell)\end{array}\right)
\label{hjt}
\end{equation}
\begin{equation}
{\cal H}_{\rm br}=-\sqrt{2}(1+\beta) g \sum_{\ell} Q_1(\ell)
[c_X^{\dagger}(\ell)c_X(\ell)+c_Y^{\dagger}(\ell)c_Y(\ell)]
\label{hbr}
\end{equation}
where the breathing amplitude is $Q_1=\sqrt{2/3}(Q_x+Q_y+Q_z)$,
$Q_2$ is $Q_x-Q_y$, and $Q_3$ is $(2Q_z-Q_x-Q_y)/\sqrt{3}$, 
in standard Van Vleck \cite{magnetism} notation.
A non-zero $\beta$ was introduced by Millis,
to represent additional charge coupling to the breathing mode.
To simplify the model, we use $\beta=0$, which makes Eqns. 
(\ref{hjt},\ref{hbr}) identical to Eqn. \ref{hep}.
 
The hopping Hamiltonian in the new basis is:

\begin{eqnarray}
{\cal H}_{\rm t}=t\sum_{\ell,\delta=x,y,z}\left(c_X^{\dagger}(\ell), 
c_Y^{\dagger}(\ell)\right)T_{\delta}
\left( \begin{array}{c} c_X(\ell \pm \hat{\delta})\\c_Y(\ell \pm \hat{\delta})
\end{array}\right)
\nonumber\\
T_x=\left( \begin{array}{rr}  \frac{2+\sqrt{3}}{4} & -\frac{1}{4} \\
             -\frac{1}{4} & \frac{2-\sqrt{3}}{4}   \end{array} \right),
T_y=\left( \begin{array}{rr}  \frac{2-\sqrt{3}}{4} & -\frac{1}{4} \\
             -\frac{1}{4} & \frac{2+\sqrt{3}}{4}   \end{array} \right),
T_z=\left( \begin{array}{rr}  \frac{1}{2} & \frac{1}{2} \\
             \frac{1}{2} & \frac{1}{2}   \end{array} \right)
\label{hnt}
\end{eqnarray}

\section{ground state solution}

We have solved this Hamiltonian for zero doping ($x=0$) in two
opposite limits: $U/t$ small, by a Hartree-Fock calculation,
and $U/t$ infinite.  
The two limits give similar answers.
The latter case seems to us more realistic
and has another advantage: since hopping is prevented by unit
occupancy of all sites, the solution does not depend on the magnetic
order.  We postpone discussing the Hartree-Fock solution until the
end of this section.

A distortion $(Q_2,Q_3)=Q(\cos\theta,\sin\theta)
\exp(i\vec{q}\cdot\vec{\ell})$ is introduced.  We 
minimize elastic and electron energy to find optimal
distortion parameters $Q,\theta$, and $\vec{q}$.  
The optimal distortion has wavevector $\vec{q}=(\pi,\pi,\pi)$.
Unless we add anharmonic or strain terms to select 
a direction in $(Q_2,Q_3)$-space, 
the energy is independent of $\theta$.  Experiment shows that the
actual ordering is $Q_2$-type, so we simply choose this distortion,
and avoid having an extra term in the Hamiltonian.

\begin{figure}
\centerline{
\psfig{figure=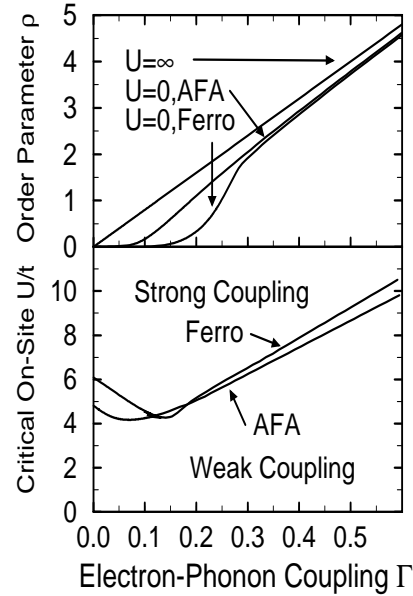,height=3.2in,width=3.0in,angle=0}}
\caption{Critical value of on-site repulsion $U$ which separates
weakly from strongly-correlated solutions.
The upper panel shows the Jahn-Teller distortion $\rho=gu/t$ as a function of
$\Gamma=g^2/Kt$ for weak and strong-coupling solutions. }
\label{ucrit}
\end{figure}

The ground electronic state which corresponds to perfect 
$Q_2$-type orbital order is
\begin{equation}
|{\rm JT}>=\prod_{\ell}^A c_X^{\dagger}(\ell)
    \prod_{\ell^{\prime}}^B c_Y^{\dagger}(\ell^{\prime})|0>,
\label{gs}
\end{equation}
where A and B label sublattices where the phase of the orbital
order $\exp(i\vec{q}\cdot\vec{\ell})$ is $\pm$1 respectively.
The energy $<{\rm JT}|{\cal H}|{\rm JT}>=-NgQ+NKQ^2/16$ 
has minimum value $E/N=-4\Gamma t$
at $Q=8g/K$ as shown on Fig. \ref{ucrit}.  
This JT phase is insulating; the gap to
charge excitations is approximately $U \approx$ 6eV. 
Electron-phonon effects are conveniently expressed in terms of the
dimensionless parameter $\Gamma=g^2/Kt$
which we estimate to be $\approx$0.25-0.35.
Orbital excitations \cite{Perebei}
require only energy $16\Gamma t \approx$2 eV, and spin excitations
occur down to low energies, with energies $\approx$50K determined by balancing
terms of order $J\approx t^2/U$.  

As explained by Goodenough \cite{Goodenough}, the $Q_2$-type orbital
order (observed below $T_{\rm JT}$=750K)
gives a layered structure (shown in Fig. \ref{jt}) which
in turn fixes the spin order which sets in below the Neel temperature
$T_N=$140K.  The magnetic structure seen \cite{Wollan} at small $x$
is ``antiferromagnetic A'' (AFA), with spins aligned ferromagnetically
within the layers and antiferromagnetically perpendicular.
The source of the ferromagnetic in-plane exchange ($J_1 S^2$=3.32 meV
\cite{Moussa}) is orbital order which favors virtual hops from
filled A sublattice $\psi_X$ orbitals to empty B sublattice
$\psi_X$ with spin parallel to avoid a Hund penalty.  When orbitals
are not ordered, the Hund penalty is outweighed by the greater multiplicity
of hops which $t_{2g}$-electrons can make onto empty $t_{2g}$-states on
antiparallel neighbors.
The antiferromagnetic c-axis exchange with $J_2 S^2$=-2.32 meV 
is explained this way.  

It turns out that none of the energetics
described so far (for $Q_2$-type orbital order) depend on the $z$-component
of the orbital ordering wave-vector $\vec{q}$, which experimentally is
$(\pi,\pi,0)$ rather than $(\pi,\pi,\pi)$ as would be preferred for any
orbital order other than pure $Q_2$.  Again, we simply adopt this order,
without specifying the additional term in the Hamiltonian.

In the opposite limit $U/t$ small, hopping energy is dominant, and
the results depend on the magnetic order.
We consider the two cases of ferromagnetic (Ferro) and AFA order.
In latter case we turn off hopping in the $z$
direction ($T_z=0$ in Eqn. \ref{hnt}.)  Again, a
distortion $(Q_2,Q_3)=Q(\cos\theta,\sin\theta)
\exp(i\vec{q}\cdot\vec{\ell})$ is introduced, and we 
minimize elastic and electron energy in Hartree-Fock (HF) approximation.
At half-filling ($x=0$) of the $e_g$-bands, 
there is perfect nesting at wavevector $\vec{q}=(\pi,\pi,\pi)$,
because of a symmetry of the bands of the undistorted
crystal, $\epsilon_1(\vec{k})=-\epsilon_2(\vec{k}+\vec{q})$.
This fixes the optimal distortion at $\vec{q}=(\pi,\pi,\pi)$.
The resulting HF energy is almost independent of the angle
$\theta$ in ($Q_2$,$Q_3$)-space, favoring the $Q_3$-distortion
by one part in $10^4$ of the JT energy.  The calculations are shown 
in Fig. \ref{ucrit}, which also shows the critical value of
$U/t$ at which the weak coupling (HF) solution and the strong-coupling
($U/t \rightarrow \infty$) solutions have equal total energy.
For LaMnO$_3$, we estimate 
$U/t \approx$ 12.  
The lower panel of Fig. \ref{ucrit} shows that
for this choice, the $U\rightarrow\infty$
approximation is better.  The upper panel shows that the JT
distortion does not depend much on $U$ for $\Gamma\ge 0.25$.

\section{small polaron}

\begin{figure}
\centerline{
\psfig{figure=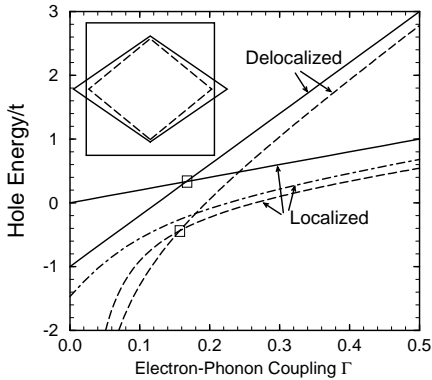,height=2.0in,width=2.3in,angle=0}}
\caption{Lowest hole energy in the AFA state of pure LaMnO$_3$
as a function of coupling constant $\Gamma$, for the trial
delocalized and localized solutions.  Solid curves are lowest order
variational solutions and dashed curves include perturbative corrections.
The dash-dot curve is an improved perturbative solution
(exact subspace diagonalization) for the localized case.
The crossover points at each level of approximation are shown in boxes.  
The inset shows
(in exaggerated form) the shape of anti-Jahn-Teller polaron (dashed lines)
in the x-y plane.  The relative
positions of O atoms (at vertices of rhombus) in the pure JT
(solid line) state and in the polaron state are accurate.}
\label{gammacr}
\end{figure}

What happens when a hole is added?  It can go into any of the 
states $|\ell A>$ and $|\ell^{\prime} B>$ which are occupied in Eqn. \ref{gs}.
When $\Gamma=0$ the hole is free to hop among these states,
no matter how large is $U$, since spins are aligned in the
planes.  When $\Gamma>0$ it costs energy $gQ=8\Gamma t$ in lost JT
energy to remove the electron.  There are two different ways
to regain some energy.  {\bf (a)} The hole can delocalize, forming
a  conducting Bloch state, with no
relaxation of the oxygen coordinates $Q({\ell})$.  For
$U=\infty$ the wavefunction in first approximation is
\begin{equation}
\psi_k=c_A\sum_{\ell}^A e^{i\vec{k}\cdot\vec{\ell}}|\ell A>
      +c_B\sum_{\ell^{\prime}}^B e^{i\vec{k}\cdot\vec{\ell}^{\prime}}
         |\ell^{\prime} B>
\label{psik}
\end{equation}
where the $N$ states $|\ell A>,|\ell^{\prime} B>$ are obtained 
from $|{\rm JT}>$ by
putting the hole on a single site; $|\ell A>=c_X(\ell)|{\rm JT}>$ and
$|\ell^{\prime} B>=c_Y(\ell^{\prime})|{\rm JT}>$.
The resulting hopping energy is
$\epsilon_k /t=(1/2)(\cos k_x + \cos k_y)$ plus an additional
term $-\cos k_z$ in the Ferro case due to hopping in $z$ direction
(prohibited in AFA case.)
The minimum energy of the extended hole state is
$E_{\rm h,ext}/t=8\Gamma - 1$ (AFA) and $8\Gamma-2$ (Ferro).
{\bf (b)} If the distortions $Q_{\ell}$ are locally readjusted,
a bound state can be formed, which will be
non-metallic because a small impurity potential will pin it.
In first approximation, put the hole at a single A site (neglecting
hopping for now.) 
The nearest $\hat{x}$-oxygens, instead of
being displaced outwards by $u=2g/K$, should now displace inwards by 
$u=-(\sqrt{4/3}-1)g/K$; the $\hat{y}$-oxygens, formerly
displaced inwards by $u=-2g/K$, now displace slightly further
inwards, by $u=-(\sqrt{4/3}+1)g/K$; the $\hat{z}$-neighbors,
formerly undisplaced, now displace in by $u=-\sqrt{4/3}g/K.$
The pattern of this anti-JT polaron is shown in Fig. \ref{gammacr}.
The energy $E_{\rm h,pol}/t$ needed to make 
the hole is reduced from
$8\Gamma$ in lost JT energy to $2\Gamma$.  This energy comes partly
from a reduction in the strain cost and partly from energy of breathing.
Setting $E_{\rm h,ext}=E_{\rm h,pol}$, we conclude that small polarons 
are stable
for $\Gamma>1/6$ in AFA phase, and that ferromagnetism, by enhancing
the delocalization energy, prevents small polarons until
$\Gamma>1/3$.  These energies are shown as the solid 
curves in Fig. \ref{gammacr}.

The three largest errors in the calculation of the hole energy 
are (1) the localized hole can spread somewhat onto neighboring
Mn atoms; (2) the Hilbert space for both localized and delocalized 
hole wavefunctions should include
states with orbital defects (sites occupied singly, but by the
orbital of wrong orientation); (3) a hole
in AFA phase will lower its energy further by causing spin
canting on Mn atoms in adjacent planes, thus permitting
interplanar hopping, and eventually driving an AFA to Ferro
transition.  We have calculated these
effects by perturbation theory.  Details for processes (1) and (2)
are given in the Appendix, and process (3) is further discussed
in the next section.

The probability that a small hole
polaron will be found away from the central site is $a/(1.0366+a)$
where $a=0.0059/\Gamma^2$ (AFA) and $0.0099/\Gamma^2$ (Ferro).
At the smallest $\Gamma$ where small polarons are stable, the
probability is 19\% and 10\%.  This shows that the small polaron is
very well localized.  The first two effects cause the
energy to be lowered by $Ct\Gamma+C^{\prime}t/\Gamma$ with
$C$=0 for delocalized states and 0.49 for localized states,
and $C^{\prime}$=(0.11,0.14) in (AFA, Ferro) for delocalized states
and (0.11, 0.17) for localized.  The shifts are small for
values of $\Gamma \ge 0.2$.  The critical value of $\Gamma$ for
small polaron formation changes from $\Gamma_c=1/6$ to 0.157 
(AFA) and from 1/3 to 0.294 (Ferro).
The corrected energies are shown as the dashed
curves in Fig. \ref{gammacr}.

\section{magnetic transition}

\begin{figure}
\centerline{
\psfig{figure=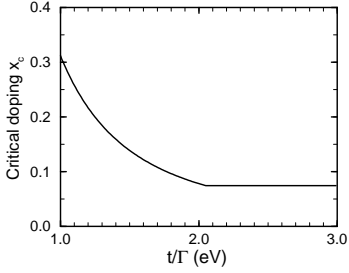,height=2.0in,width=2.3in,angle=0}}
\caption{Critical concentration $x_c$ for the AFA to Ferro transition
as a function of coupling constants $t/\Gamma$.}
\label{xcrit}
\end{figure}

Effect (3), the spin-canting problem, has usually been discussed
on the assumption that holes are delocalized \cite{deGennes}.
Treating spins classically,
the additional delocalization energy gained when adjacent
layers realign from 180$^{\circ}$ to 180$^{\circ} -\theta$ 
is $t\sin(\theta/2)$ per hole; the exchange cost of realignment
is $J_2 S^2 (1-\cos(\theta))$ per Mn atom.  The optimum
canting angle is $\sin(\theta/2)=xt/4|J_2|S^2$, which gives
a small critical concentration $x_c=4|J_2|S^2 /t$ for complete
rotation to the Ferro phase.  Localized holes tilt spins on
only near-neighbor Mn atoms in adjacent planes, gaining less
hopping energy than delocalized holes
because the electron hops into an anti-JT-distorted hole 
site.  The energy gain is $(39/640)(t/\Gamma)\sin(\theta/2)$.
If we neglect the rotation of any except first-neighbor spins, the 
magnetic energy loss around the localized hole is
$(8J_1 +2|J_2|)S^2 (1-\cos(\theta))$ per hole, giving the
optimum local rotation $\sin(\theta/2)=0.49$(eV$^{-1})t/\Gamma$.  
For $t>2.04\Gamma$ eV, the adjacent spins are completely flipped.
Comparing the magnetic energy loss per spin with the energy loss
of the Ferro state, we find a critical concentration plotted
in Fig. \ref{xcrit}.  Experimentally the AFA/Ferro phase
boundary occurs at $x_c$=0.08 for Sr-doped LaMnO$_3$ \cite{Kawano}
and at $x_c$=0.15 for Ca doping \cite{Wollan}.  
These values are easily reconciled
with the localized hole picture, but demand too small
a value of $t$ in the delocalized picture.  The smaller Ca ion
causes larger Mn-O-Mn bond angles which results in a smaller $t$
and a larger $x_c$.

\section{metal-insulator transition}

At doping concentrations $x$ approaching $x_c$=0.20, LaMnO$_3$
becomes metallic.  A possible analog system with simpler but
similar physics is doped BaBiO$_3$.  When Ba atoms are replaced
by K atoms, the system remains insulating up to a concentration
$x$=0.40.  The most likely and successful explanation for the 
insulating behavior is formation of bi-polarons.  The pure
material has Bi atoms in the nominal Bi$^{4+}$ valence state.
Chemically Bi prefers valences Bi$^{3+}$ and Bi$^{5+}$.
The pure material has 
alternating oxygen displacements inward or
outward in the $Q_1$ breathing pattern.  This 
stabilizes charge ordering which is interpretable as 
alternation of Bi$^{3+}$ and Bi$^{5+}$.  When K atoms
replace Ba, extra electrons are released from Bi atoms,
creating holes, and  
converting some of the Bi$^{3+}$ ions into Bi$^{5+}$. 
The new Bi$^{5+}$ ions created by doping are small hole bi-polarons,
and are stabilized by relaxation of the $Q_1$ coordinates of oxygens.
A microscopic description has been given \cite{Kostur} using
the Rice-Sneddon Hamiltonian \cite{Rice}, which is just a simplified
version of the Hamiltonian used here, with a single Bi $s$-band
in place of the two Mn $e_g$-bands.  The Hubbard U is much smaller
for Bi $s$ orbitals than for Mn $e_g$'s, 
so bi-polarons are the stable defect
of the Peierls-type charge order parameter, while 
single polarons are the stable defect of
the JT orbital order parameter of LaMnO$_3$.  

The metal-insulator transition in BaBiO$_3$ was addressed in
a remarkable calculation by Yu, Chen, and Su \cite{Yu} (verified
by Kostur and Allen \cite{unpublished}).  In their calculations,
bipolaron defects form spontaneously in the Peierls order parameter
up to very large concentrations, but when the concentration is
too high, the defects destroy the Peierls order completely,
and the ground state switches from defective dimerized insulator to
undistorted metal.  In the work of Kostur and Allen \cite{unpublished}
it is observed that the defective dimerized state (a ``bipolaron
glass'' phase) is probably an Anderson insulator.  The bipolarons
are dense enough that localized single-particle states fill
up the Peierls gap, and an Efros-Shklovskii pseudogap \cite{Efros}
rather than a clean gap separates the
filled and empty states at the Fermi level.
The insulating state might have a non-zero linear specific
heat coefficient $\gamma$.
However, the metal-insulator transition is not an Anderson
transition -- the localized states do not persist away from
the Fermi level in the metallic state, but disappear altogether.

The transition in LaMnO$_3$ seems to us to be similar,
with polarons instead of bipolarons, and JT rather
than Peierls order being destroyed at the
phase transition.  We have not attempted a calculation, which
would be more difficult for LaMnO$_3$ than for BaBiO$_3$
because of the large value of $U$.  Schematically the energy
of JT glass phase and metallic phase would be something like
\begin{eqnarray}
E_{\rm JT}/Nt &=& -[4-a(U)x-b(U)x^2+\cdots]\Gamma \nonumber \\
E_{\rm metal}/Nt &=& -[c(U)x + d(U)x^2 +\cdots]
\label{eschem}
\end{eqnarray}
where coefficients $a,b,c,d$, etc. are unknown functions which depend
strongly on $U$ and also on the magnetic state.  Our calculations above
give the value of $a$ at $U=\infty$ to be 2 plus small corrections.
If we could neglect $bx$ relative to $a$ and $dx$ relative to $c$,
then the critical concentration would be $x_c=4\Gamma/(a\Gamma+c)$.  The
observed $x_c \approx 0.2$ then requires that the hopping energy gain $-ct$
in the correlated metal phase must be of order -$5t$, using $\Gamma=0.3$,
whereas we find that at $U=0$ the value of $c$ in the
ferromagnetic phase is only 3.  This shows
that the polaron repulsion $b$ plays a significant role.

\section{discussion}

Numerous effects are left out
of the model.  Macroscopic strain, tilting of oxygen octahedra,
and additional zero-point magnetic and non-adiabatic
lattice fluctuations all could be
added and would affect the $T=0$ properties discussed here, but we
believe that the effects we did include are the most important ones.
A $T>0$ treatment of all these effects is a bigger challenge.
It would also be interesting to extend the $T=0$ 
calculations to higher doping where a remarkable variety of
textured phases is being unraveled \cite{Bao}.
The properties of LaMnO$_3$-related materials are incredibly rich, 
yet surprisingly understandable, in contrast to
certain other transition-metal oxide systems.
 
\acknowledgements
We thank W. Bao, M. Blume, D. E. Cox, J. P. Franck, J. B. Goodenough, 
J. P. Hill, D. I. Khomskii, and 
J.-S. Zhou for helpful conversations, and V. N. Kostur for the inspiration to
start the project.
This work was supported in part by NSF grant no. DMR-9725037.

\appendix
\section*{perturbative corrections}

In Sec. IV the energy of the hole in JT-distorted LaMnO$_3$
is computed in lowest order,
for the two cases of localized and delocalized hole states.
Here the method used for perturbative corrections is explained.
First we discuss the localized case.

Denoting the pure crystal ground state Eqn. \ref{gs} as $|{\rm JT}>$,
the strictly localized hole on an A site at the origin is
$c_X(0)|{\rm JT}>$.  For clarity we use both a shorthand
and a pictorial labeling scheme:
\begin{equation}
|h0>=c_X(0)|{\rm JT}>=\left|\begin{array}{ccc}
& \v &\\ \v &{\sf O}& \v  \\ & \v & \end{array}\right\rangle
\label{h0}
\end{equation}
where $h0$ means ``hole at the origin''.  This is the
zeroth order (unperturbed) state of the localized hole.
As shown in section IV, relaxation of oxygen positions reduces the
energy $\epsilon(h0)=<h0|{\cal H}_{\rm tot}|h0>$ from $8\Gamma t$ to
$2\Gamma t$.  Our aim is to calculate the two corrections,
\begin{equation}
\delta E_1 = \sum_{i=1}^6 \frac{\left|\left<i|{\cal H}_{\rm JT}
               |h0\right>\right|^2}
             {\epsilon(h0)-\epsilon(i)} 
\label{dele1}
\end{equation}
\begin{equation}
\delta E_2 = \sum_{j=1}^{12} \frac{\left|\left<j|{\cal H}_{t}
               |h0\right>\right|^2}
             {\epsilon(h0)-\epsilon(j)} ,
\label{dele2}
\end{equation}
Included in the sums are all basis functions which couple to
$|h0>$ by either ${\cal H}_{\rm JT}$ or ${\cal H}_{t}$.
All our basis
states have the same relaxed oxygen positions optimized
for the $|h0>$ state.

The JT operator ${\cal H}_{\rm JT}$ couples $|h0>$ to six
states $|i>$ with orbital defects on nearest neighbor Mn atoms.
The states $|i>$ come in three types.  First consider
\begin{equation}
|h0,o\hat{x}>=c_X^{\dagger}(\hat{x})c_Y(\hat{x})c_X(0)|{\rm JT}>
=\left|\begin{array}{ccc}
& \v &\\  \v &{\sf O}& \h \\ & \v & \end{array}\right\rangle
\label{h0ox}
\end{equation}
and the related state $|h0,o-\hat{x}>$.
The label $|h\hat{s},o\hat{s}^{\prime}>$ has the
meaning ``hole at site $\hat{s}$, orbital defect at site $\hat{s}^{\prime}$.''
When there is no orbital defect, the second part of the label is omitted.
These states have energy $\epsilon(h0,o\pm\hat{x})-\epsilon(h0)=
\left(14-4/\sqrt{3}\right)\Gamma t$, 
reduced from the value $16\Gamma t$ of
a distant orbital defect by relaxation of the displaced
oxygen lying between the origin and the $\pm\hat{x}$ Mn.  
These states are coupled to $|h0>$ in first
order by the ${\cal H}_{JT}$ term proportional to
$Q_3(\pm\hat{x})=-\left(2/3+1/\sqrt{3}\right)g/K$.  
Second are the states $|h0,o\pm\hat{y}>$ with the hole 
at the origin and the orbital defect at the $\pm\hat{y}$ sites.
These have a slightly increased energy denominator 
$\left(14+4/\sqrt{3}\right)\Gamma t$, and coupling 
amplitude to the $|h0>$ state caused by the oxygen relaxation
$Q_3(\pm\hat{y})=-\left(2/3-1/\sqrt{3}\right)g/K$.
Third, the states $|h0,o\pm \hat{z}>$ have the same energy denominator
$16\Gamma t$ as a distant orbiton, because the only neighboring oxygen with
an altered position lies along the $\hat{z}$ axis, generating the
term $Q_3(\pm\hat{z})=4g/3K$ which causes off-diagonal coupling.
Adding the three types of correction, we get the value for
the first perturbative correction to the localized hole energy,
\begin{equation}
\delta E_1=-\frac{628}{1287}\Gamma t\approx-0.488 \Gamma t.
\label{E1}
\end{equation}
This correction is independent of magnetic ordering.

The hopping Hamiltonian ${\cal H}_t$ couples state $|h0>$
to twelve states, which come in two categories of six each.
One example of each category is:
\begin{equation}
|h\hat{x}> = c_Y(\hat{x})|{\rm JT}>=\left|\begin{array}{ccc}
& \v &\\  \v & \h &{\sf O} \\ & \v & \end{array}\right\rangle
\label{hx}
\end{equation}
\begin{equation}
|h\hat{x},o0>=c_Y^{\dagger}(0)c_X(0)c_Y(\hat{x})|{\rm JT}>
    =\left|\begin{array}{ccc}
& \ \v &\\  \v & \ \v &{\sf O} \\ & \ \v & \end{array}\right\rangle
\label{hxo0}
\end{equation}
The first six have the hole displaced to a first neighbor site, with
the origin occupied by an electron in the ``correct'' orbital; the 
second six have a misoriented orbital at the origin.
The hops without orbital defect in the $\pm(\hat{x},\hat{y})$ directions cost
energy $\epsilon(j)-\epsilon(h0)=37 \Gamma t/3$, 
while the hops in the $\pm\hat{z}$ direction (forbidden in the AFA
magnetic state) cost $40 \Gamma t/3$.  Moving a  hole to a remote
site would cost slightly less energy $12 \Gamma t$.
Hops which leave an orbital defect at the origin cost an extra $8\Gamma t$.
The coupling magnitude is determined by the 
off-diagonal part of the matrix of Eqn. \ref{hnt} for hopping without creating 
orbital defect at the origin and by the diagonal parts otherwise.
The net result is
\begin{eqnarray}
\delta E_2({\rm Ferro}) &=& -\left(\frac{3}{148}+\frac{3}{80}+
       \frac{21}{244}+\frac{3}{128}
        \right)\frac{t}{\Gamma}\approx -0.167\frac{t}{\Gamma} \nonumber\\
\delta E_2({\rm AFA}) &=& -\left(\frac{3}{148}+\frac{21}{244}
       \right)\frac{t}{\Gamma}\approx -0.106\frac{t}{\Gamma}.
\label{E2}
\end{eqnarray}

These calculated shifts are shown in the AFA case as a dashed line
in Fig. \ref{gammacr}.  The shifts are similar in magnitude to the
unperturbed energy.  Therefore we repeated the calculation with an
exact diagonalization in the same subset of 18 states which couple
to $|h0>$.  The answer, shown as the dot-dashed line
in Fig. \ref{gammacr}, is not greatly different from the perturbative
answer in the relevant regime of parameters.

In section IV the case of the delocalized hole was treated
by staying within the subspace of states $c(\ell)|{\rm JT}>$
where $c(\ell)$ removes an occupied JT orbital, and then
diagonalizing the off-diagonal effects of ${\cal H}_t$ by
Fourier transformation to Bloch states.  The hole is then
put into the lowest energy Bloch state.  This calculation has
omitted effects caused by the fact that ${\cal H}_t$ also
allows hops which leave behind a single orbital defect.
We now correct for this perturbatively.  Our Hilbert space
has two subspaces.  The Bloch states lie in the $N$-dimensional
subspace with a single hole and no orbital defect.  We must add
a $6N$-dimensional space in which the hole has an orbital defect
on an adjacent atom.  These two subspaces are coupled by ${\cal H}_t$.
The Schr\"odinger equation has the structure
\begin{equation} \left(
\begin{array}{cc} {\cal H}_{\rm JT}+{\cal H}_t -E & {\cal H}_t^{\prime} \\
                  {\cal H}_t^{\prime}  &  {\cal H}_{\rm JT}  -E \end{array} 
                  \right)
\left( \begin{array}{c} \psi_{\rm I} \\ \psi_{\rm II} \end{array} \right)=0.
\end{equation}
\label{schdeloc}
Each element of the $\psi_{\rm I}$ subspace (no orbital defect) is
coupled to six elements of the $\psi_{\rm II}$ subspace by hopping
terms.  The prime on ${\cal H}_t^{\prime}$ is used to designate the
part of ${\cal H}_t$ which creates an orbital defect.
The JT energy is constant ($8\Gamma t$ hole creation energy)
in the $\psi_{\rm I}$ subspace  and higher by the JT gap ($16\Gamma t$)
in the $\psi_{\rm II}$ subspace.  In our perturbative treatment
we leave out the off-diagonal effects of ${\cal H}_t$ interior to
the $\psi_{\rm II}$ subspace.  The problem is then equivalent to
an effective Hamiltonian
\begin{equation}
{\cal H}_{\rm eff}={\cal H}_{\rm JT}+{\cal H}_t
-{\cal H}_t^{\prime} ({\cal H}_{\rm JT}  -E)^{-1} {\cal H}_t^{\prime}
\label{heff}
\end{equation}
in the $\psi_{\rm I}$ subspace.  Second-order perturbation
theory uses the JT gap $16\Gamma t$ as the energy denominator
$({\cal H}_{\rm JT}  -E)$.  Then the last term in Eqn. \ref{heff}
just gives a constant shift on the diagonal.  The value depends
on the magnetic state.
\begin{eqnarray}
\delta E_3({\rm Ferro}) & = & -\left(\frac{7}{64}+\frac{2}{64}
\right)\frac{t}{\Gamma}\approx -0.141\frac{t}{\Gamma}
\nonumber\\
\delta E_3({\rm AFA}) & = & -\frac{7}{64}\frac{t}{\Gamma}
\approx -0.109\frac{t}{\Gamma}
\label{E3}
\end{eqnarray}
This shift is shown as a dashed line (in the AFA case) in Fig.
\ref{gammacr}.


\end{document}